\documentclass[journal]{IEEEtran}

\usepackage{multirow}
\usepackage{color}
\usepackage{cite}
\ifCLASSINFOpdf
\usepackage[pdftex]{graphicx}
\graphicspath{{figure/}}
\DeclareGraphicsExtensions{.pdf,.jpeg,.jpg,}
\else
\fi
\ifCLASSOPTIONcompsoc
 \usepackage[caption=false,font=normalsize,labelfont=sf,textfont=sf]{subfig}
\else
  \usepackage[caption=false,font=footnotesize]{subfig}
\fi
\begin{document}
\title{Reconstruction for Diverging-Wave Imaging \\ Using Deep Convolutional Neural Networks}
\author{Jingfeng~Lu,
        Fabien~Millioz,
        Damien~Garcia,
        S{\'e}bastien~Salles,
        Wanyu~Liu,
        and~Denis~Friboulet
\thanks{J. Lu is with METISLab, School of Instrumentation Science and Engineering, Harbin Institute of Technology, Harbin, China, and also with University of Lyon, CREATIS, CNRS UMR 5220, Inserm U1044, INSA-Lyon, University of Lyon 1, Villeurbanne, France (email: lujingfeng@hit.edu.cn).}
\thanks{F. Millioz, D. Garcia, S. Salles, and D. Friboulet are with the University of Lyon, CREATIS, CNRS UMR 5220, Inserm U1044, INSA-Lyon, University of Lyon 1, Villeurbanne, France.}
\thanks{W. Liu is with METISLab, School of Instrumentation Science and Engineering, Harbin Institute of Technology, Harbin, China, and also with Sino European School of Technology of Shanghai University, Shanghai, China.}
}
\maketitle
\begin{abstract}
In recent years, diverging wave (DW) ultrasound imaging has become a very promising methodology for cardiovascular imaging due to its high temporal resolution. However, if they are limited in number, DW transmits provide lower image quality compared with classical focused schemes. A conventional reconstruction approach consists in summing series of ultrasound signals coherently, at the expense of frame rate, data volume, and computation time. To deal with this limitation, we propose a convolutional neural networks (CNN) architecture, IDNet, for high-quality reconstruction of DW ultrasound images using a small number of transmissions. In order to cope with the specificities induced by the sectorial geometry associated to DW imaging, we adopted the inception model composed of the concatenation of multi-scale convolutional kernels. Incorporating inception modules aims at capturing different image features with multi-scale receptive fields. A mapping between low-quality images and corresponding high-quality compounded reconstruction was learned by training the network using \textit{in vitro} and \textit{in vivo} samples. The performance of the proposed approach was evaluated in terms of contrast ratio (CR), contrast-to-noise ratio (CNR) and lateral resolution (LR), and compared with standard compounding method and conventional CNN methods. The results demonstrated that our method could produce high-quality images using only 3 DWs, yielding an image quality equivalent to that obtained with compounding of 31 DWs and outperforming more conventional CNN architectures in terms of complexity, inference time and image quality. 
\end{abstract}


\IEEEpeerreviewmaketitle

\section{Introduction}

\IEEEPARstart{U}{ltrasound} imaging has become the modality of choice for cardiovascular imaging because of its noninvasive, cost-efficient, and real time properties. In conventional ultrasound imaging, several sectors of the entire image are reconstructed using sequential narrow beams. The frame rate of conventional schemes mainly depends on the number of transmitted beams required to construct an image. Limitations arise when monitoring highly transient biological phenomena faster than the frame rate delivered by the conventional scheme. Tracking mechanical waves, such as remotely induced shear waves \cite{couade2010vivo} and electromechanical waves \cite{provost2009electromechanical}, is a representative example of such challenge as their propagation speed in myocardium can reach 1 to 10 m/s \cite{papadacci2014high}.

To reduce the number of transmissions, multi-line acquisition \cite{shattuck1984explososcan} and multi-line transmit \cite{mallart1992improved} have been introduced. Multi-line acquisition approach uses broader transmit beam and reconstruct multiple image lines (e.g., 2, 4, or 8) for each transmission. In the multi-line transmit method, multiple focused beams are simultaneously transmitted. In order to increase frame rate while preserving the number of scan lines, ultrafast imaging using unfocused transmit beam has been proposed. These approaches use plane waves (PW) \cite{sandrin1999time} or diverging waves (DW) \cite{hasegawa2011high, poree2016high} to image a wide field of view. However, in PW or DW imaging, the acoustic energy of unfocused beams is spread onto a wider area, resulting in a deterioration of the quality of reconstructed images if no additional processing is performed. 

To alleviate these effects, coherent compounding \cite{montaldo2009coherent,tanter2014ultrafast,poree2016high} of unfocused beams has been proposed. Coherent compounding consists in transmitting multiple consecutive beams at different angles. The backscattered echoes are then coherently summed to improve contrast and resolution. Therefore, a trade-off needs to be made between image quality and frame rate since compounding of more beams produces images of higher quality but decreases the image rate. Therefore, to achieve a high image quality while maintaining the frame rate of original unfocused wave imaging is of high interest. 

In recent years, deep learning has achieved state-of-the-art performance in various problems of image processing, including image classification, recognition, and segmentation. The success of deep learning methods lies in its exponentially increasing expressiveness, which can capture modality-specific features \cite{lecun2015deep}. Inspired by the success of deep learning, many researchers have investigated deep learning methods for medical image reconstruction and achieved significant performance \cite{jin2017deep, wolterink2017generative, han2018framing, kang2018deep, wang2016accelerating, schlemper2017deep, hammernik2018learning, lee2018deep}. For instance, Jin et al. \cite{jin2017deep} proposed to use convolutional neural networks (CNN) to generate high-quality X-ray computed tomography (CT) images from low-quality images reconstructed from sparse views. Lee et al. \cite{lee2018deep} proposed a deep residual learning network \cite{he2016deep} for the reconstruction of magnetic resonance (MR) images from accelerated MR acquisition. 

Regarding ultrasound imaging, a number of studies have recently been carried out on image reconstruction based on deep learning \cite{gasse2017high, perdios2018deep, hyun2019beamforming, khan2019deep, luchies2018deep, luijten2019deep, nair2018deep, senouf2018high, vedula2019learning, vedula2018high, yoon2018efficient, zhang2018high, zhou2018high, zhuang2019deep, lu2019fast, ghani2019high}. Among these works, only a few studies were devoted to the specific problem of PW reconstruction \cite{perdios2018deep, khan2019deep, luijten2019deep, nair2018deep, zhang2018high, zhou2018high}. The studies in \cite{luijten2019deep, nair2018deep} dealt with PW reconstruction as a beamforming problem (i.e., raw radio frequency (RF) data were processed) and thus did not consider the possibility of synthetic focusing by compounding multiple PWs. Other studies \cite{perdios2018deep, khan2019deep, zhang2018high, zhou2018high} processed post-beamformed data, possibly taking advantage of several PW transmissions for compounding. Our group previously proposed to reconstruct high-quality images using 3 PW transmissions \cite{gasse2017high}. A compounding operation was learned using a fully convolutional network. We demonstrated that this approach could produce high-quality images using 3 PWs while preserving the image quality close to that obtained by standard compounding with 31 PWs. In \cite{zhang2018high}, Zhang et al. used a generative adversarial networks (GAN) where the generator was mainly composed of 50 residual blocks, each of them consisting of two convolutional layers. As in \cite{gasse2017high}, the reconstruction was performed from 3 PWs, and the results were found by the authors to be similar to those of \cite{gasse2017high} in terms of contrast-to-noise-ratio (CNR) and lateral resolution (LR). In \cite{zhou2018high}, Zhou et al. proposed to use a 3-scales CNN with feedback, where the processing of a scale used up to 5 convolutional layers. However, as the proposed CNN did not preserve speckle texture, they resorted to a complex, non-linear, wavelet-based post-processing step to improve the reconstruction. The obtained results were found slightly better than \cite{gasse2017high} (difference of 1.8 dB in terms of peak signal-to-noise-ratio (PSNR), 0.02 in terms of Structural Similarity Index (SSIM) and 0.08 in terms of Mutual Information (MI)). Khan et al. described a CNN consisting of 27 convolution layers in a contracting path with concatenation \cite{khan2019deep}. The CNR of the reconstruction obtained with 3 PWs (2.56 dB) was found to be approximately equivalent to that of the standard delay-and-sum (DAS) reconstruction obtained with 11 PWs (2.51 dB). Perdios et al. proposed to use a straightforward U-Net with 5 decomposition levels \cite{perdios2018deep}. The reconstruction was evaluated from simulations and yielded very encouraging results, i.e., a CNR of 17.1 dB and a lateral resolution of 0.38 mm in the far field.

PW and DW have been shown to be suitable for high-frame-rate imaging in echocardiography \cite{cikes2014ultrafast}. They both allow a significant decrease in transmit numbers to examine the region of interest. However, as with focused sequences, echocardiographic PW imaging requires a series of sequential beams that scan the area from edge to edge. In contrast, each DW transmit can cover the entire region of interest. The DW technique has thus two advantages: i) a further gain in image rate; ii) the different regions of the heart are imaged synchronously. As a result, DW-based echocardiography allows simultaneous analysis of intracardiac flow and myocardial motion, as shown by Faurie et al. \cite{faurie2016intracardiac}. Using a DW transmission scheme, it is also possible to cover a wide and deep cardiac view (a four-chamber view) at a very high frame rate (up to 400 cardiac-images/s). Taking advantage of this feature, the motion of the entire myocardium can be deciphered accurately by standard \cite{joos2018high} or advanced \cite{poree2018dual} methods of speckle tracking. The problem of reconstructing images from DW transmissions using deep learning techniques has been addressed only very recently in \cite{lu2019fast}, which was a preliminary version of the present paper, and by Ghani et al. in \cite{ghani2019high}. The latter study used a six-layer fully convolutional neural network operating patch-wisely from 11 transmissions. As we will show in the following, the CNN architecture we proposed used only 3 DWs and provided an image quality that competed with the standard compounding of 31 DWs. It is difficult to draw further conclusions from the comparison with \cite{ghani2019high} to the extent that this study only provided qualitative results (i.e., images).

In this paper, we introduce a novel CNN architecture for high-quality reconstruction for DW imaging using a small number of DW transmissions. To achieve this goal, a unique issue must be resolved. Conventional CNN architectures adopt fixed kernels in one convolutional layer where the same weights are applied over the entire feature map. Such shared-weight architecture contributes to achieving the shift-invariant feature of CNN \cite{zhang1990parallel}, which is well-adapted to PW images. Nevertheless, due to the sectorial geometry induced by DW acquisition, applying deep learning approaches to DW images is specific. Processing RF lines obtained from DW acquisitions implicitly means that the CNN operates in polar coordinates. Maintaining the shift-invariance feature of CNN methods (that was directly applicable in the Cartesian coordinate system associated with PW imaging) requires to adapt the CNN architecture.
Inspired by the GoogleNet \cite{szegedy2015going} for image classification problems, we therefore incorporated inception modules to the fully convolutional architecture for the reconstruction of DW images, yielding the so-called IDNet (Inception for DW Network). Inception modules employed convolution filters of different sizes for the same input and concatenate all the output for the next layer. As it will be shown in the sequel, inception used in conjunction with maxout activation allowed features from multiple receptive field sizes to be captured, in contrast to conventional CNN architectures where the receptive field size was fixed. Therefore, different image features of different image regions can be learned via multiple convolution filters. This property is demonstrated in more detail in the discussion section.

In summary, the contributions of this work are the following:

1) We address the problem of reconstruction for DW imaging using deep learning.

2) We introduce a CNN architecture with the inception module to take into account the specific geometry of DW imaging.

3) We demonstrate that, using only 3 DWs, our method yielded high-quality images competing with those obtained by compounding with 31 DWs.

4) We further show that the proposed method could work at a high frame rate, which made it amenable to real-time reconstruction for DW imaging. 

The remainder of this paper is organized as follows: in Section II, the proposed method is described. Section III presents the details of data acquisition and implementation for training. Section IV presents the experiment results that validate the effectiveness of the proposed method. We further discuss the proposed method in Section V and conclude the work in section VI.

\begin{figure*}[!t]
\centering
\includegraphics[width=5.5in]{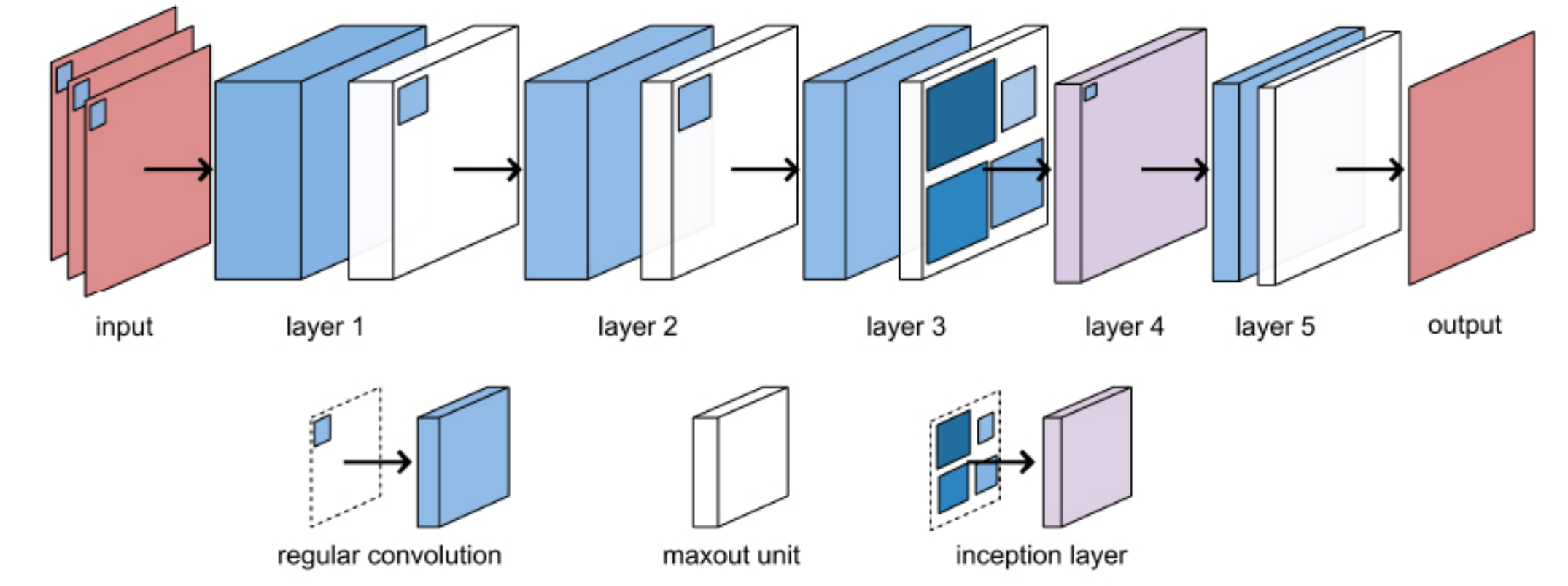}
\caption{Block diagram of the architecture of the proposed network.}
\label{network architecture}
\end{figure*}

\begin{table*}[!t]
\renewcommand{\arraystretch}{1.5}
\caption{Architecture of proposed network}
\label{network}
\centering
\begin{tabular}{c c c c c c}
\hline
\multirow{2} * {block type} & feature size & kernel size & padding & number of & activation\\ 
~ & channel $\times$ height $\times$ width & height $\times$ width & height $\times$ width & kernels  & - \\
\hline
inputs & $m \times h \times w$ & - & - & - & -\\
\hline
convolution & $64 \times h \times w$ & $9 \times 3$ & $4 \times 1$ & 256 & maxout 4\\
\hline
convolution & $32 \times h \times w$ & $17 \times 5$ & $8 \times 2$ & 128 & maxout 4\\
\hline
convolution & $16 \times h \times w$ & $33 \times 9$ & $16 \times 4$ & 64 & maxout 4\\
\hline
\multirow{4} * {inception} &\multirow{4} * {$8 \times h \times w$} & $41 \times 11$ & $20 \times 5$ & 8 & maxout 4\\
~ & ~ & $49 \times 13$ & $24 \times 6$ & 8 & maxout 4\\
~ & ~ & $57 \times 15$ & $28 \times 7$ & 8 & maxout 4\\
~ & ~ & $65 \times 17$ & $32 \times 8$ & 8 & maxout 4\\
\hline
convolution & $1 \times h \times w$ & $1 \times 1$ & - & 4 & maxout 4\\
\hline
\end{tabular}
\end{table*}

\section{Methods}
\subsection{Problem formulation}
Let $X$ be a tensor that contains the low-quality RF beamformed images of size $m \times w \times h$, where $m$ is the number of DW transmissions, $w$ is the number of scan lines, and $h$ is the length of each RF signal. Our method aimed at producing one reconstructed RF image of size $w\times h$ using the input $X$. Standard compounding consists of summing all $m$ DWs to obtain the high-quality image. Considering that there might be useful information which was not exploited by standard compounding, we employed a CNN with trainable parameters $\Theta$ to learn the optimal mapping $f(\cdot)$ of $X \rightarrow Y$, where $Y$ was the reference obtained from the standard compounding of $n$ ($n \gg m$) DWs.

\subsection{Network architecture}
Fig. \ref{network architecture} is a pictorial description of the proposed IDNet architecture. IDNet was a 2-D convolutional network composed of 5 hidden layers. Two types of basic building modules were employed to build the network. In Fig. \ref{network architecture}, each blue block denotes a regular convolutional module followed by a maxout unit activation (white block), and the purple block indicates the inception module. The choices and details related to this architecture are discussed hereunder.

\textit{Fully convolution architecture}. We excluded the pooling operation from both the regular convolution layers and the inception layer to produce the feature maps with the same dimension. This guaranteed that the spatial information was preserved at the same scale throughout the network, which was beneficial for maintaining phase in RF signals. The kernel size of each layer was doubled compared to that of its previous layer to achieve the effect of doubling receptive field size, the same as the 2 $\times$ 2 pooling operation.

\begin{figure}[!t]
\centering
\includegraphics[width=3.3in]{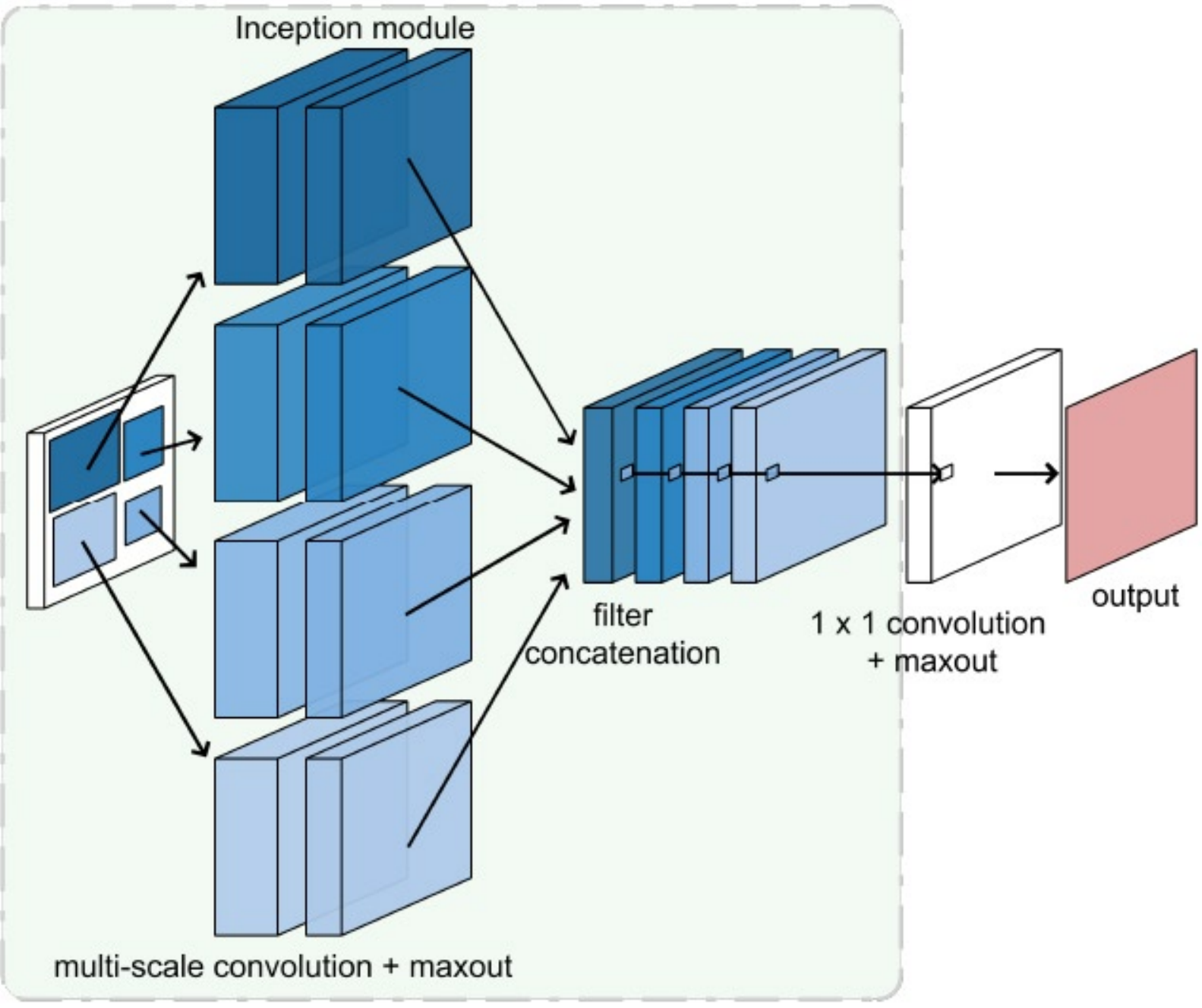}
\caption{Block diagram of the inception module and the one-by-one convolution.}
\label{inception}
\end{figure}

\begin{figure}[!t]
\centering
\includegraphics[width=3.3in]{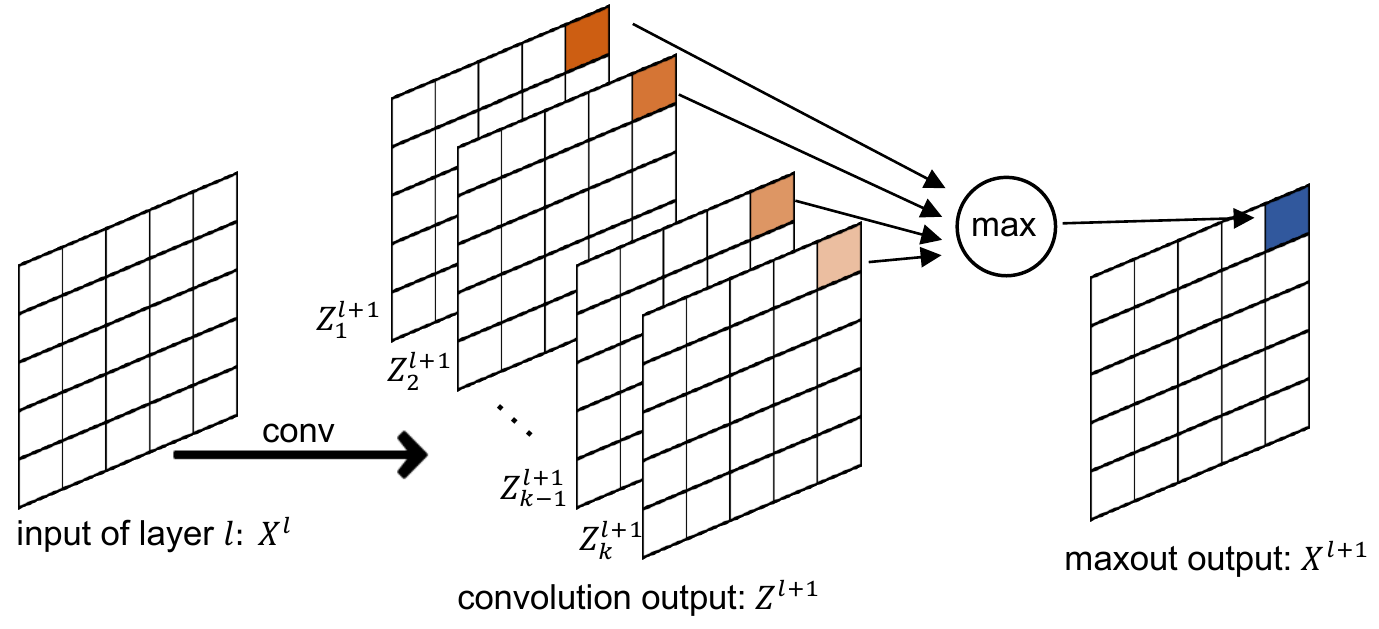}
\caption{Block diagram of a $k$ pieces maxout unit. A $k$ pieces maxout unit takes the pixel-wise maximum values across the $k$ affine feature maps ($Z_1^{l+1}$, $Z_1^{l+1}$, ..., $Z_1^{l+1}$) produced from the convolution.}
\label{maxout block}
\end{figure}

\textit{Inception module.} As illustrated in Fig. \ref{inception}, the inception module used in our network consisted of four parallel paths. Each path performed convolution of a different kernel size to the same input feature maps. All the paths performed proper padding to ensure their outputs had the same size. The outputs of all paths were stacked along the feature dimension as the final output of the inception module. As the image properties of DW imaging varied along image depth, the multi-size convolution kernels contributed to extract different image features from multiple receptive field sizes. The inception module was only used in the second last layer and followed by the 1 $\times$ 1 convolution layer for two reasons: i) in our network, the number of feature maps in shallow layers were much larger than the deep layers. Incorporating inception modules to shallow layers would drastically increase the number of parameters, since an inception module with four parallel paths would quadruple the parameters of a regular convolution; ii) As deep layers produced high-level features and large receptive sizes, more relevant information could be perceived by employing the inception module with larger kernels in the deeper layers.

\textit{One-by-one convolution.} The last layer of the network was a $1 \times 1$ convolution layer followed by a maxout unit. It served two purposes: i) to reduce the number of output channels and generate the final image; ii) All input pixels of the same position collapsed to one output pixel via the $1 \times 1$ convolution in conjunction with the maxout unit, which acted as selecting gates for every element of the inputs. The network was thus trained to select the main elements contributing to form the output element, and thus to learn different image features for different image positions.

\textit{Maxout unit.} We used maxout units \cite{goodfellow2013maxout} as the activation function for both the regular convolution layers and the inception layer. Maxout units are piecewise-linear convex functions, and a maxout network with more than two maxout units can approximate many popular activation functions and most often outperform them \cite{gasse2017high, zhao2017improving}. In a convolution network, a maxout unit takes the pixel-wise maximum values across several affine feature maps to achieve a nonlinear transformation. Fig. \ref{maxout block} is an illustration of the $k$ pieces maxout unit. Given an input $X^{l}$ to the $l$th hidden layer of a CNN, the output $X^{l+1}$ after the convolution and the maxout unit is defined as follows:
\begin{equation}
\label{maxout}
X^{l+1} = max\lbrace Z_1^{l+1}, ..., Z_k^{l+1}  \rbrace
\end{equation}
where $Z_j^{l+1}$ ($j\in[1,k]$) is the $j$th group of feature maps obtained by performing convolution to $X^{l}$, and $k$ is the number of maxout pieces.

A more detailed specification of the network architecture, such as the number of feature maps and the size of convolution kernels, is provided in Table \ref{network}. 

\section{Experiment}

\subsection{Data Set Acquisition }

A Verasonics system research scanner (Vantage 256) equipped with an ATL P4-2 probe (bandwidth: 2-4 MHz, center frequency: 3 MHz, and cycle number: 1.5) was used to perform steered DW acquisitions. The acquisition planes were acquired by continuously moving the probe on the surface of the imaged objects, at an imaging rate of 50 fps and a packet size of 250 images. Each acquisition plane was acquired using 31 steered DWs with transmission angles ranging between $\pm$ 30$^{\circ}$, with an incremental step of 2$^{\circ}$. For each DW acquisition, the received RF signals were sampled at 12 MHz and beamformed with the conventional DAS method. Each RF beamformed image was of dimension  512$\times$256, covering a sectorial region of size 11 cm $\times$ 90$^{\circ} $. The input images $X$ were composed of a small subset of $m$ = 3 DWs (-30$^{\circ}$, 0$^{\circ}$, and 30$^{\circ}$), while the reference images $Y$ were the standard compounding of all $n$ = 31 DWs. A total of 7000 ($X,Y$) samples (i.e., acquisition pairs) were used in the experiment. Specifically, 1500 acquisitions were performed on \textit{in vivo} tissues (thigh muscle, finger phalanx, and liver regions), and 5500 acquisitions were performed on \textit{in vitro} phantoms.

\subsection{Network Training}

5000 ($X,Y$ samples were randomly selected from the entire data set as the training set, 1000 ($X,Y$) samples were used as an independent validation set, and the remaining 1000 ($X,Y$) samples were used as the testing set for evaluation.
 Learning the reconstruction mapping function $f(\cdot)$ required the estimation of the optimal network parameters $\Theta$ by minimizing the loss between the reconstructed images $\hat Y = f(X;\Theta)$ and the reference $Y$. Mean Squared Error (MSE) was used as the loss function:
\begin{equation}
\label{mse}
L(\Theta) = \frac{1}{n} \sum_{i=1}^n \| f(X_i;\Theta) - Y_i \|^2,
\end{equation}
where $n$ is the number of training samples and $i \in [1,n]$ denotes the index of each training sample.

In the training stage, the network weights were initialized with the Xavier initializer \cite{glorot2010understand}. The loss was minimized using mini-batch gradient descent with the Adam optimizer \cite{kingma2014adam}, and the batch size was set to 10. The initial learning rate was set to $1 \times 10^{-4}$ and an early stopping strategy was employed to adjust the learning rate. The learning rate was halved if there had been no decrease in the validation loss for 20 epochs, and 40 epochs without validation loss reduction would end the training. The training was performed using Pytorch \cite{paszke2017automatic} library on a NVIDIA Tesla V100 GPU with 32 Gb of memory, resulting in training time of about two days.

\subsection{Evaluation Metrics}

To quantitatively show the effectiveness of the proposed network, we used peak-signal-to-noise ratio (PSNR), structure similarity (SSIM), mutual information (MI), contrast ratio (CR), contrast-to-noise ratio (CNR), and lateral resolution (LR) as the evaluation metrics. PSNR and SSIM were used to evaluate the reconstruction quality by comparing it to the reference images (i.e., images obtained through the standard compounding of 31 DWs).  The other three quantities (CR, CNR, and LR) were used as specialized ultrasound indices to evaluate the cyst images and the width of the point spread function from point target images.

PSNR  is defined as the ratio of the maximum possible power of a signal and the distorting noise which deteriorates the quality of its representation. Given the MSE between the reconstructed image and the reference, the PSNR is calculated as follows:
\begin{equation}
\label{psnr}
PSNR = 20 \log_{10}\frac{MAX_I}{\sqrt{MSE}},
\end{equation}
where $MAX_I$ is the max pixel value of the image.

SSIM measures the similarity between two images, and MI measures mutual dependence between two images. Given the reconstructed image $\hat{Y}$ and the reference $Y$, the SSIM and MI are calculated as follows:
\begin{equation}
\label{ssim}
SSIM = \frac{(2\mu _{\hat Y} \mu _{Y} + C_1)(2\sigma _{\hat{Y}{Y}} + C_2)}{(\mu _{\hat Y}^2 + \mu _Y^2 + C_1)(\sigma _{\hat Y}^2 + \sigma _Y^2 + C_2)},
\end{equation}

\begin{equation}
\label{mi}
MI = \sum_{\hat{y},y}p_{\hat{Y}Y}(\hat{y},y)log\frac{p_{\hat{Y}Y}(\hat{y},y)}{p_{\hat Y}(\hat{y})p_Y(y)}
\end{equation}
where $\mu _{\hat Y}$ and $\mu _Y$ ($\sigma^2_{\hat Y}$ and $\sigma^2_Y$) denote the means (variances) of $\hat Y$ and $Y$, $\sigma _{\hat{Y}{Y}}$ denotes the covariance between $\hat Y$ and $Y$, $C_1$ and $C_2$ are two constants that stabilize the division with weak denominator, $p_{\hat{Y}Y}(\hat{y},y)$ is the joint distribution of $\hat Y$ and $Y$, and $p_{\hat Y}(\hat y)$ and $p_Y(y)$ are the marginal probability distribution of $\hat Y$ and $Y$.
 
CR and CNR were used to measure the contrast between the object of interest and the surrounding background:
\begin{equation}
\label{cr_formula}
CR = -20 \log_{10}\frac{ \mu _t}{\mu _b},
\end{equation}
\begin{equation}
\label{cnr_formula}
CNR = 20 \log_{10}\frac{\vert \mu _t - \mu _b\vert}{\sqrt{\sigma^2_t + \sigma^2_b}},
\end{equation}
where $\mu _t$ and $\mu _b$ ($\sigma^2_t$ and $\sigma^2_b$) denote the means (variances) of the intensity within the target region and the background. 

LR was used to assess the width of the point spread function from point target images. The full width at half maximum was used in this work to estimate the LR. 

In the testing phase, the PSNR, SSIM, and MI were computed from the full set of testing samples. The CR and CNR were measured on two anechoic regions (in the near field at 40 mm depth and the far field at 120 mm depth) of an image obtained from the Gammex phantom. The LR was measured on 0.1 mm Nylon monofilaments (in the near field at 20 mm and 40 mm depth, the middle field at 60 mm and 80 mm depth, and the far field at 90 mm and 100 mm depth) of an image obtained from the CIRS phantom. 

\begin{table}[!t]
\renewcommand{\arraystretch}{1.5}
\caption{Architectures of the inception layers of IDNet-2, IDNet-4, and IDNet-8}
\label{comparison}
\centering
\begin{tabular}{c c c c}
\hline
\multirow{2} * {model} & number of & kernel size & number of\\ 
~ & kernel types & height $\times$ width & kernels \\
\hline
\multirow{2} * {IDNet-2} &\multirow{2} * {2} & $49 \times 13$ & 16\\
~ & ~ & $65 \times 17$ & 16\\
\hline
\multirow{4} * {IDNet-4} &\multirow{4} * {4} & $41 \times 11$ & 8\\
~ & ~ & $49 \times 13$ & 8\\
~ & ~ & $57 \times 15$ & 8\\
~ & ~ & $65 \times 17$ & 8\\
\hline
\multirow{8} * {IDNet-8} &\multirow{8} * {8} & $37 \times 11$ & 4\\
~ & ~ & $41 \times 11$ & 4\\
~ & ~ & $45 \times 13$ & 4\\
~ & ~ & $49 \times 13$ & 4\\
~ & ~ & $53 \times 15$ & 4\\
~ & ~ & $57 \times 15$ & 4\\
~ & ~ & $61 \times 17$ & 4\\
~ & ~ & $65 \times 17$ & 4\\
\hline
\end{tabular}
\end{table}

\subsection{Inception and activation layer parameters}

In the proposed network, we used the inception module to exploit features with multiple receptive fields of the images. To demonstrate the effect of the inception module of our network, models with different inception layers were trained with the same training data and implementation. Each model employed an inception layer with different convolution kernels. For the sake of clarity, these models were named IDNet-2, IDNet-4, and IDNet-8, whose inception structures are shown in Table \ref{comparison}. For a fair comparison, the other components of the network shared the same architecture. Besides, to verify the effectiveness of maxout unit activation, we conducted another experiment which consisted in replacing the maxout unit with the ReLU activation (IDNet-ReLU ).

\begin{figure*}[!t]
\centering
\subfloat[]{\includegraphics[width=2.4in]{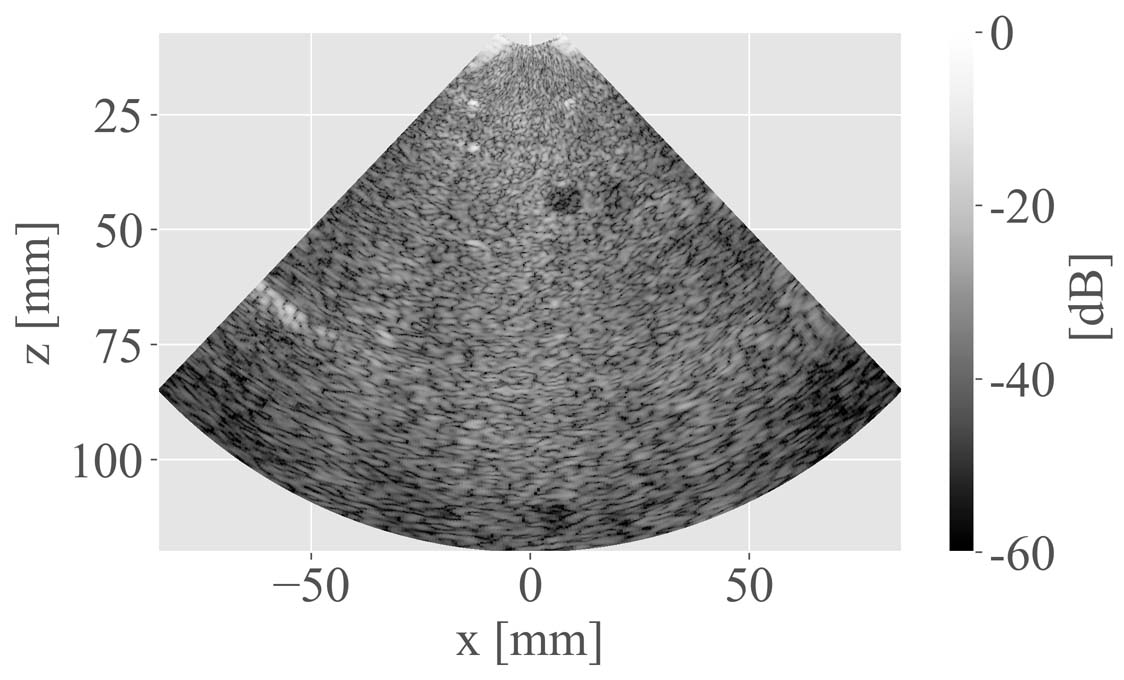}%
\label{crinput}}
\subfloat[]{\includegraphics[width=2.4in]{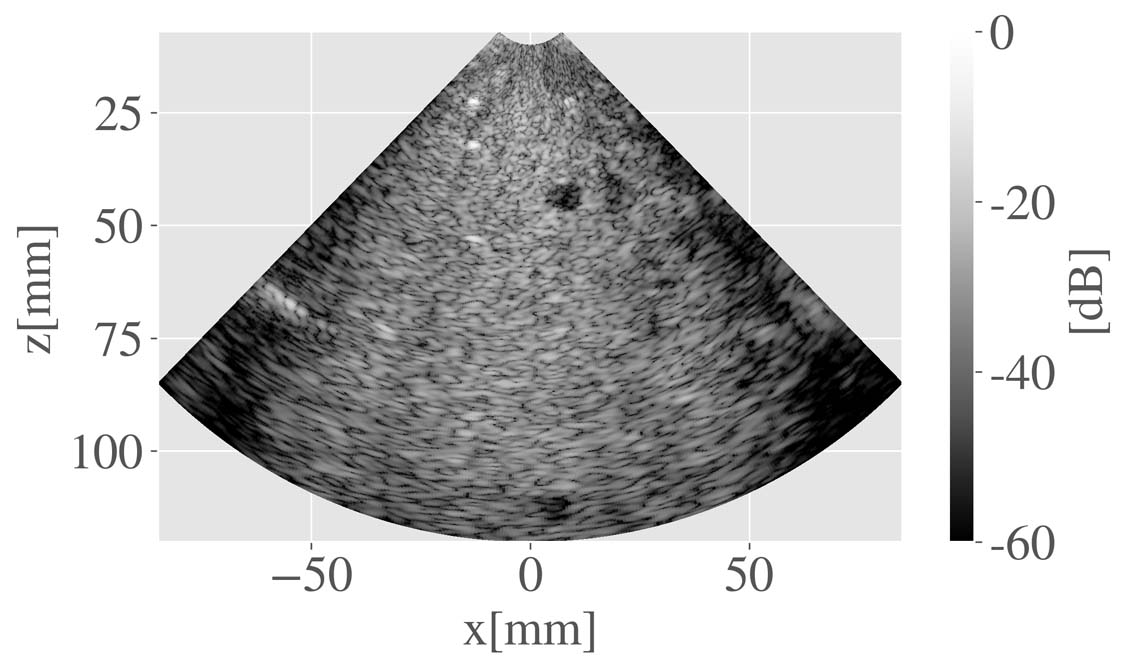}%
\label{crIDNet2}}
\subfloat[]{\includegraphics[width=2.4in]{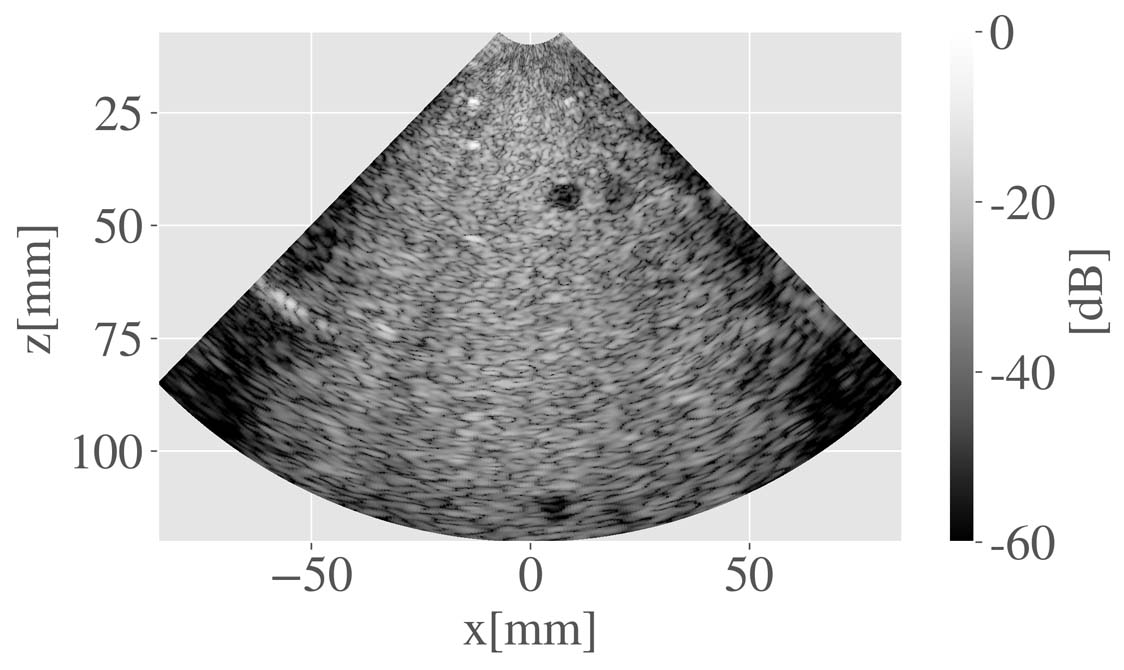}%
\label{crIDNet4}}
\hfil
\subfloat[]{\includegraphics[width=2.4in]{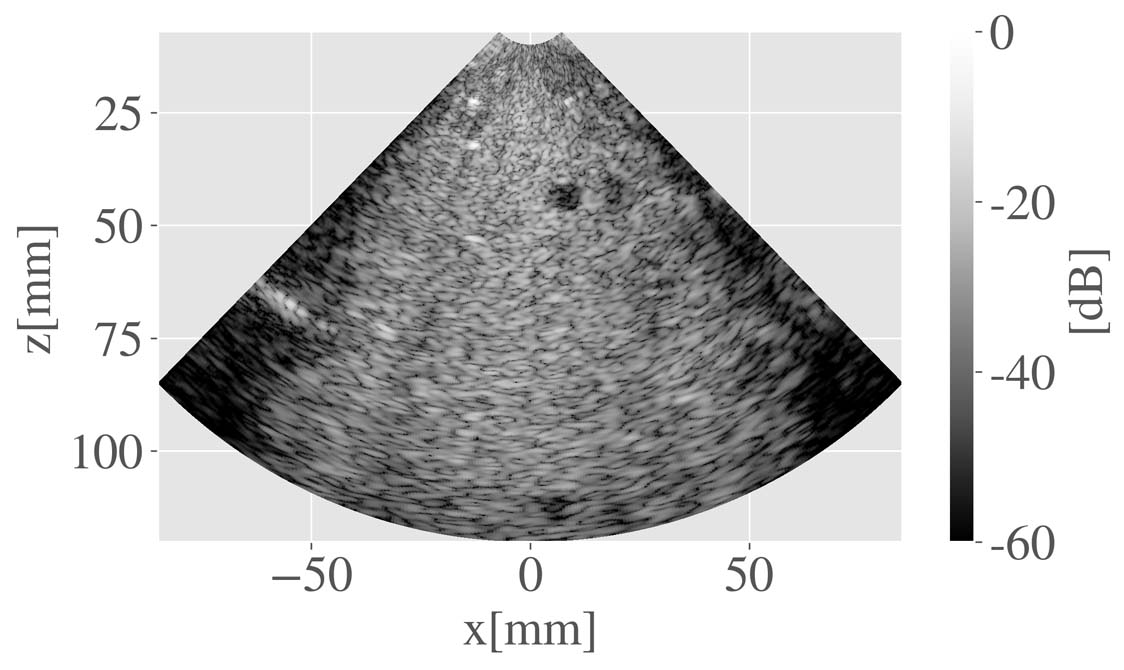}%
\label{crIDNet8}}
\subfloat[]{\includegraphics[width=2.4in]{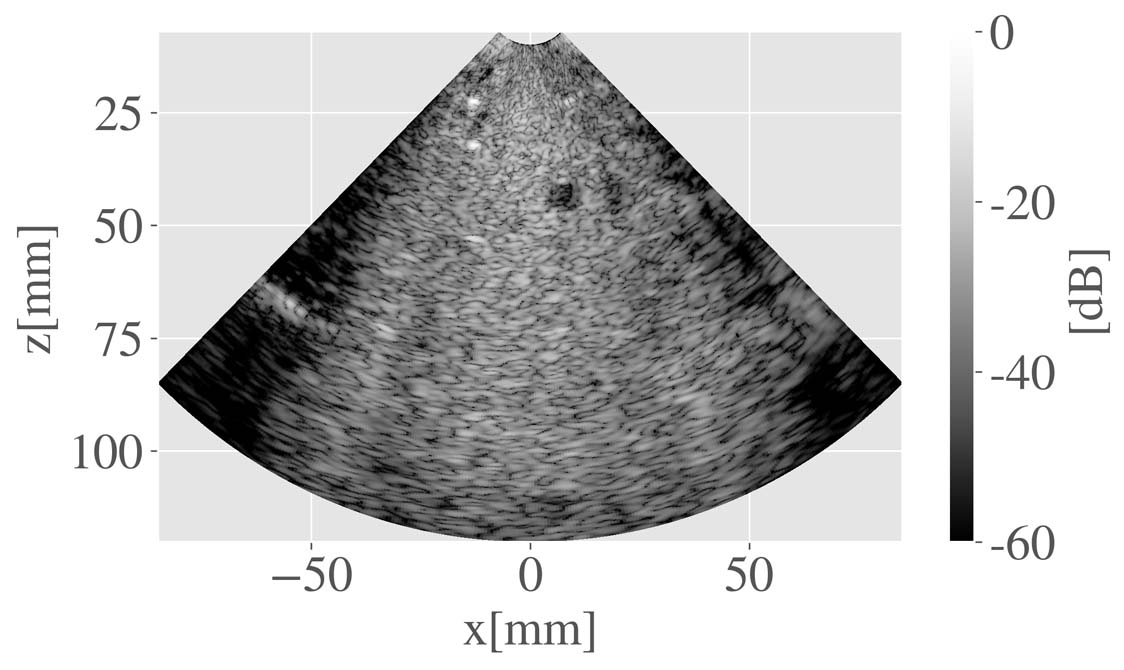}%
\label{crIDNetrelu}}
\subfloat[]{\includegraphics[width=2.4in]{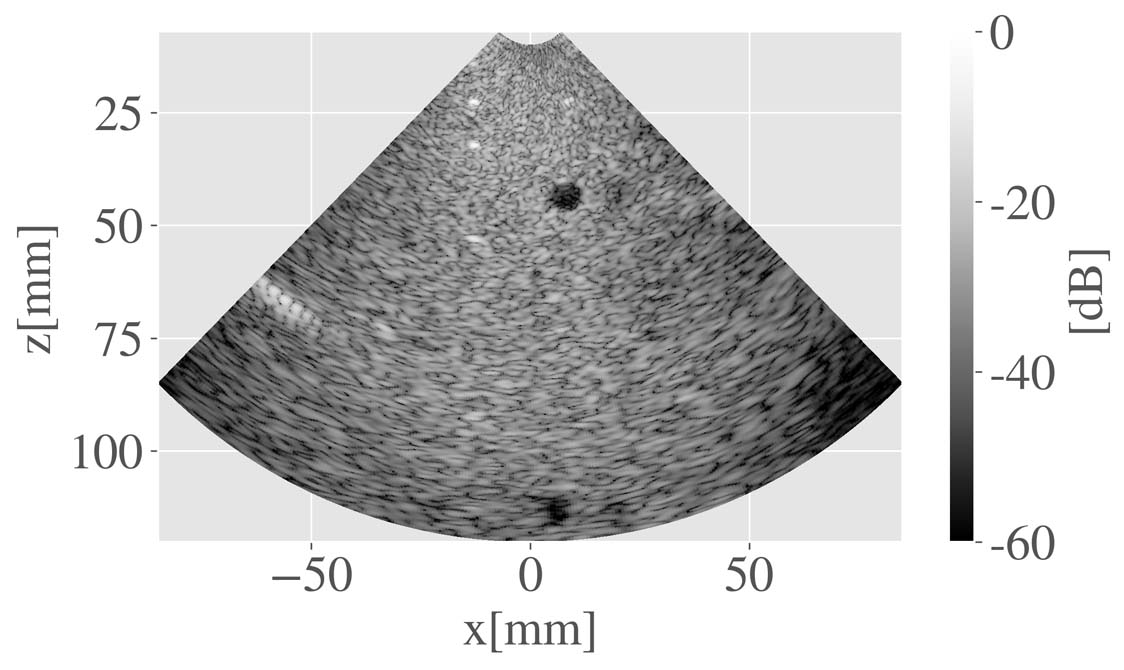}%
\label{crreference}}
\caption{Example of B-mode images reconstructed from (a) standard compounding of 3 DWs, (b) IDNet-2, (c) IDNet-4, (d) IDNet-8, (e) IDNet-ReLU, and (f) standard compounding of 31 DWs (reference).}
\label{crIDNetfigures}
\end{figure*}

\begin{table*}[!t]
\renewcommand{\arraystretch}{1.6}
\caption{Evaluation metrics of IDNet and compounding method}
\label{inception test}
\centering
\begin{tabular}{c c c c c c c c c c c}
\hline
\multirow{2} * {model} & \multirow{2} * {PSNR [dB]} &  \multirow{2} * {SSIM} & \multirow{2} * {MI} & \multicolumn{2}{c}{CR [dB]}  & \multicolumn{2}{c}{CNR [dB]} & \multicolumn{3}{c}{LR [mm]} \\ 
\cline{5-11}
~ & ~ & ~&~ & near field & far field & near field & far field & near field & middle field & far field \\
\hline
IDNet-2  & 31.06 $\pm$ 1.49  & 0.92 $\pm$ 0.06 & 0.81 $\pm$ 0.19 & 19.20 & 13.71 & 7.25 & 3.94 & 0.95 & 1.67 & 2.47\\

IDNet-4 & \textbf{31.13 $\pm$ 1.47} & \textbf{0.93 $\pm$ 0.06} & 0.82 $\pm$ 0.20 & \textbf{19.54} & \textbf{14.95} & \textbf{7.63} & \textbf{5.21} & \textbf{0.90} & 1.64 & \textbf{2.35} \\

IDNet-8 & 31.07 $\pm$ 1.50 & \textbf{0.93 $\pm$ 0.06} & \textbf{0.83 $\pm$ 0.20} & 19.05  & 13.24 & 7.41  & 4.83 & 0.96 & \textbf{1.61} & 2.44\\

IDNet-ReLU & 30.96 $\pm$ 1.50 & 0.92 $\pm$ 0.07 & 0.76 $\pm$ 0.20 & 17.04 & 11.86 & 5.37  & 2.79 & 1.01 & 1.70 & 2.48\\
Reference & - & - & - & 19.97 & 15.45 & 7.71 & 5.33 & 0.94 & 1.63 & 2.42\\

\hline
\end{tabular}
\end{table*}

\subsection{Comparison Methods}

The proposed method was compared with three methods for the evaluation of the reconstruction quality. 

1) Standard compounding method \cite{tanter2014ultrafast}: To assess the improvement of our method over standard compounding method, same input images were used to obtain the compounding reconstruction. 

2) CNN of Gasse et al. \cite{gasse2017high} 
: IDNet and Gasse's CNN shared the same architecture in the first three convolutional layers, i.e., fully convolutional layers followed by maxout activation without spatial pooling.  The difference was that we employed the inception layer composed of four parallel multi-scale convolutions, rather than the fixed convolution. Each path of the inception layer had 8 channels, and stacking of all channels produced 32 output channels, the same as the last layer in Gasse's network.

3) U-Net \cite{perdios2018deep}: U-Net is a typical encoder-decoder structure consisting of a symmetric downsampling and upsampling path. We experimentally observed that the direct use of the architecture proposed in \cite{perdios2018deep} gave poor results. To obtain a better comparison between the U-Net and our network, the convolutional sizes of the U-Net were modified to fit the data used in our experiment. The sequential operation [$3 \times 3$ convolution, ReLU, $3 \times 3$ convolution, ReLU] used in \cite{perdios2018deep} was replaced by the [$3 \times 1$ convolution, ReLU, $7 \times 3$ convolution, ReLU] operation, resulting in the same receptive field size as the $9 \times 3$ convolution of our network. 

All compared networks were trained using the dataset and settings described in the previous section.

\begin{figure*}[!t]
\centering
\subfloat[]{\includegraphics[width=2.4in]{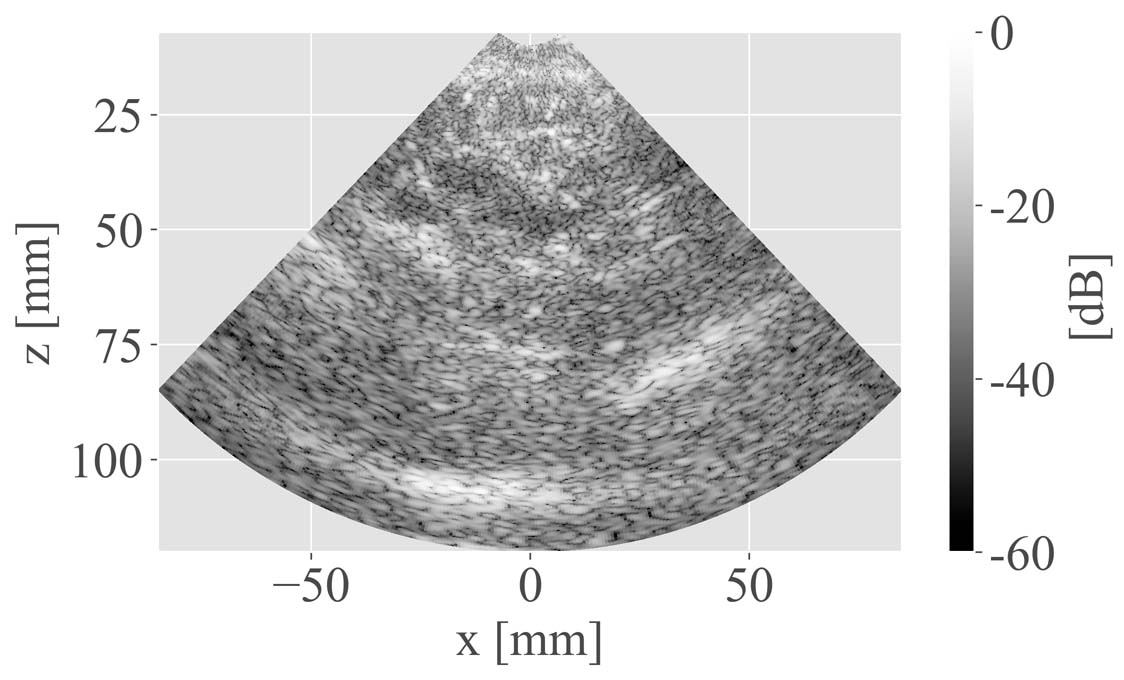}%
\label{image1_1}}
\subfloat[]{\includegraphics[width=2.4in]{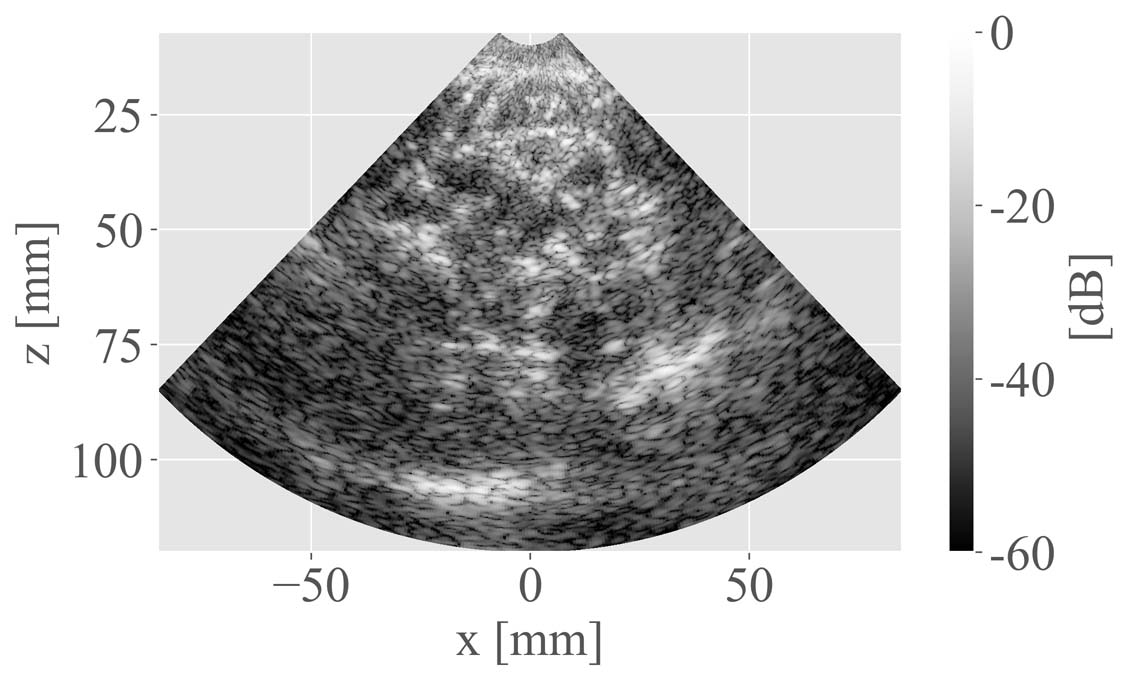}%
\label{image1_2}}
\subfloat[]{\includegraphics[width=2.4in]{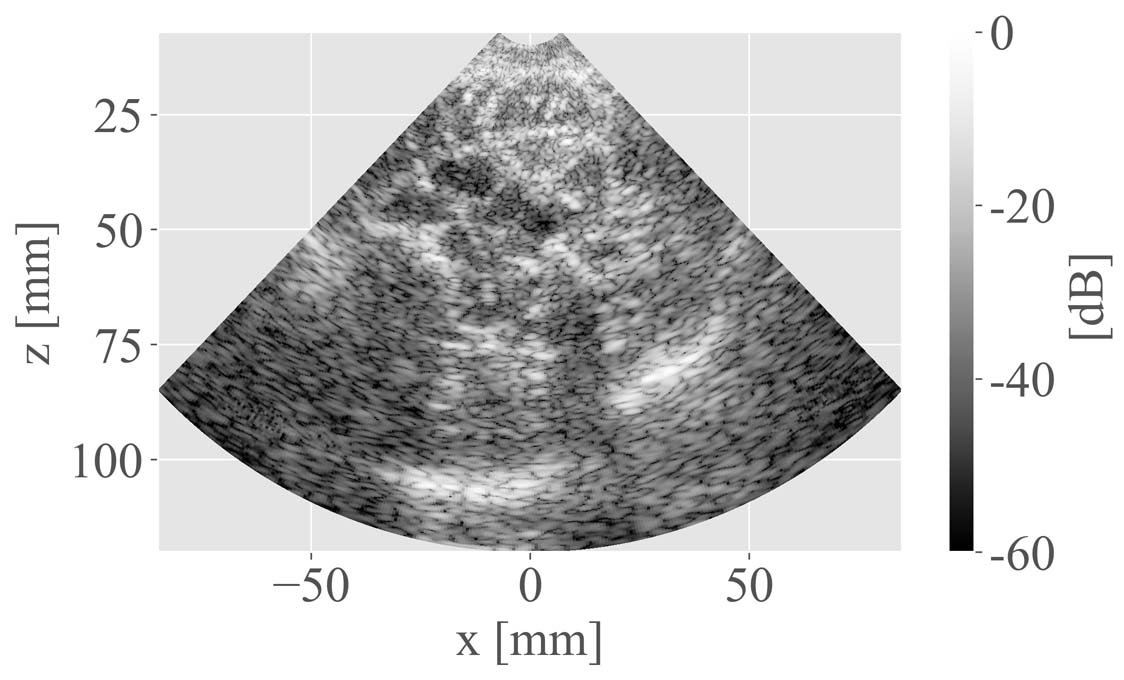}%
\label{image1_3}}
\hfil
\subfloat[]{\includegraphics[width=2.4in]{Bmode_image2_3dw.jpg}
\label{image2_1}}
\subfloat[]{\includegraphics[width=2.4in]{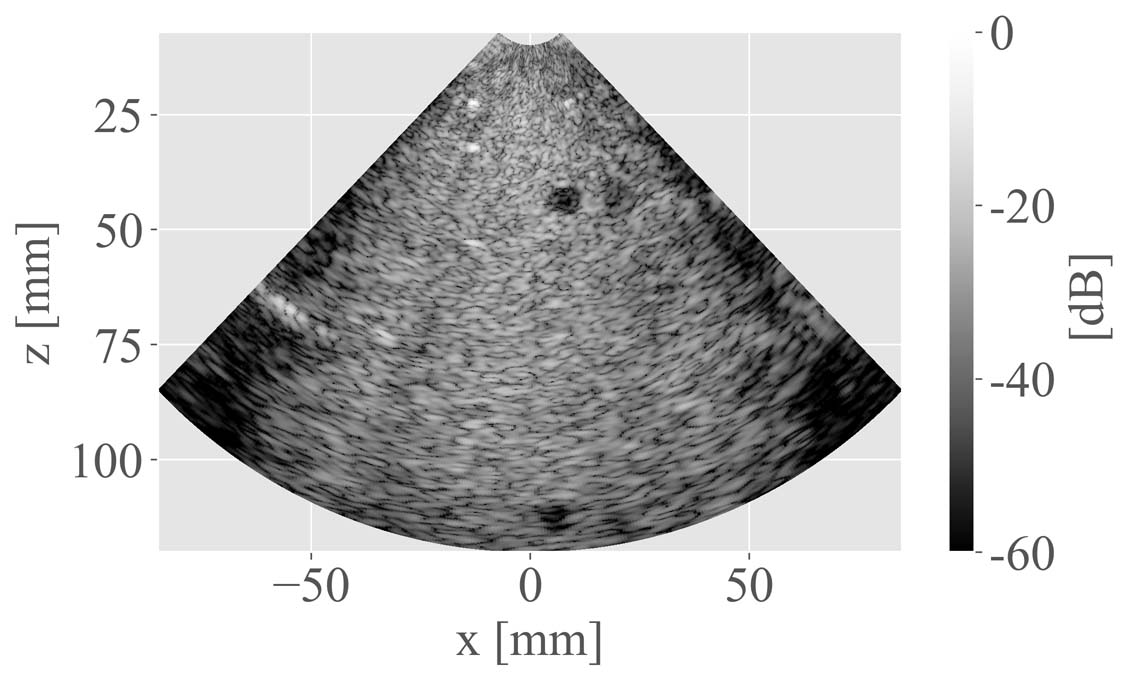}
\label{image2_2}}
\subfloat[]{\includegraphics[width=2.4in]{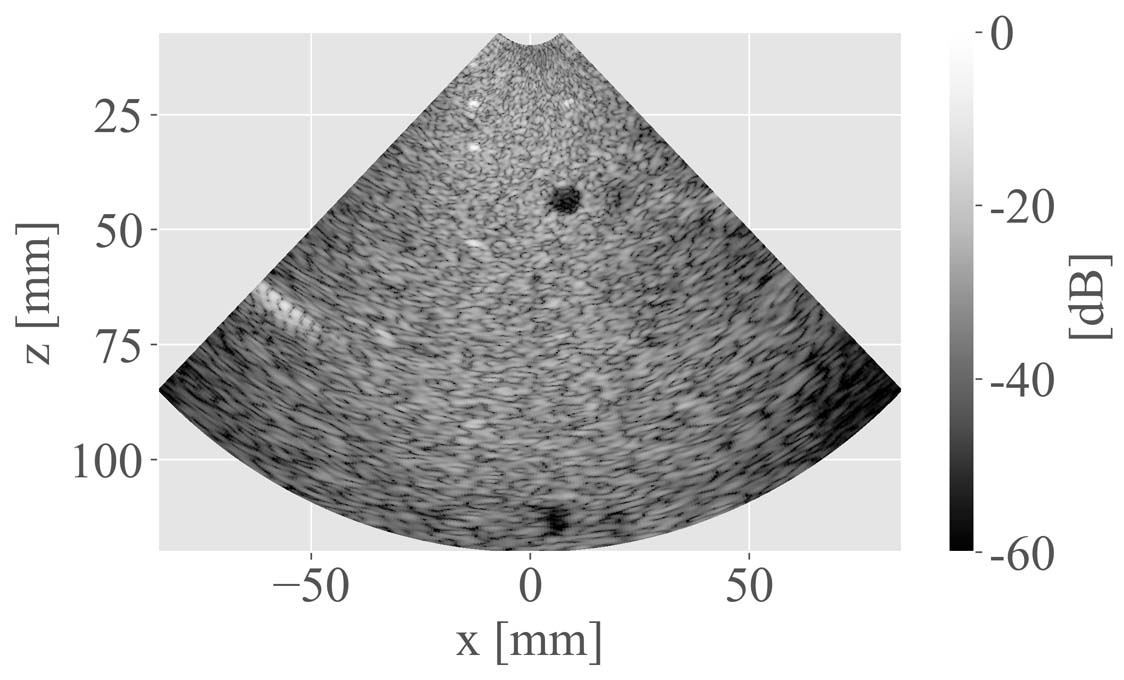}%
\label{image2_3}}
\hfil
\subfloat[]{\includegraphics[width=2.4in]{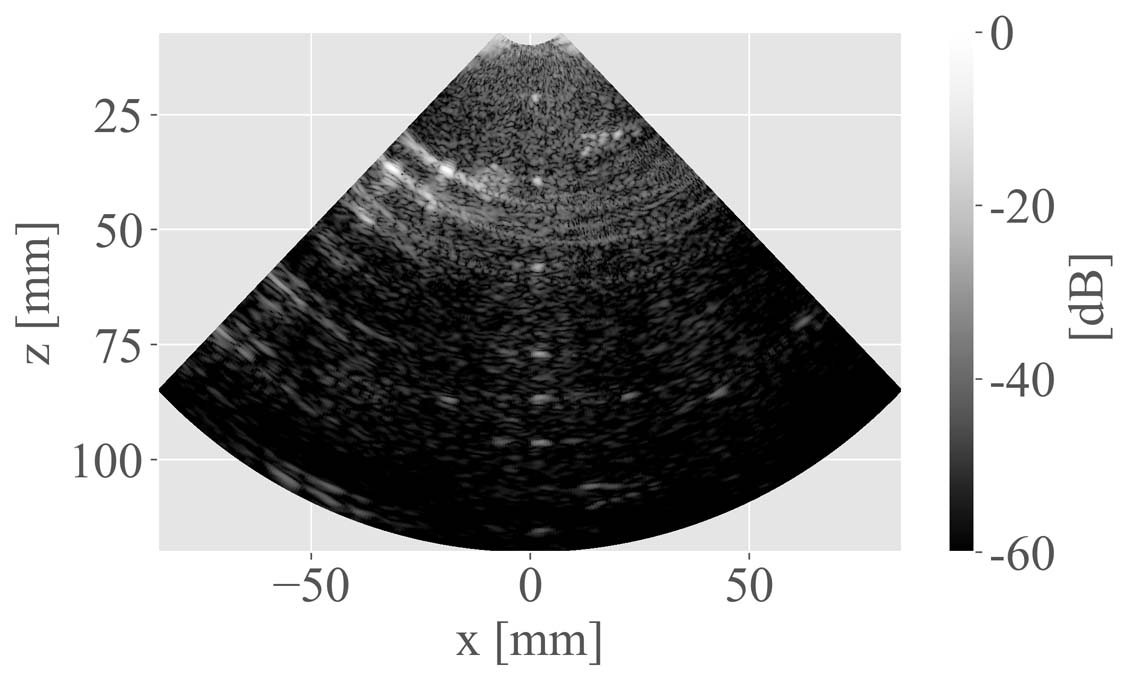}
\label{image3_1}}
\subfloat[]{\includegraphics[width=2.4in]{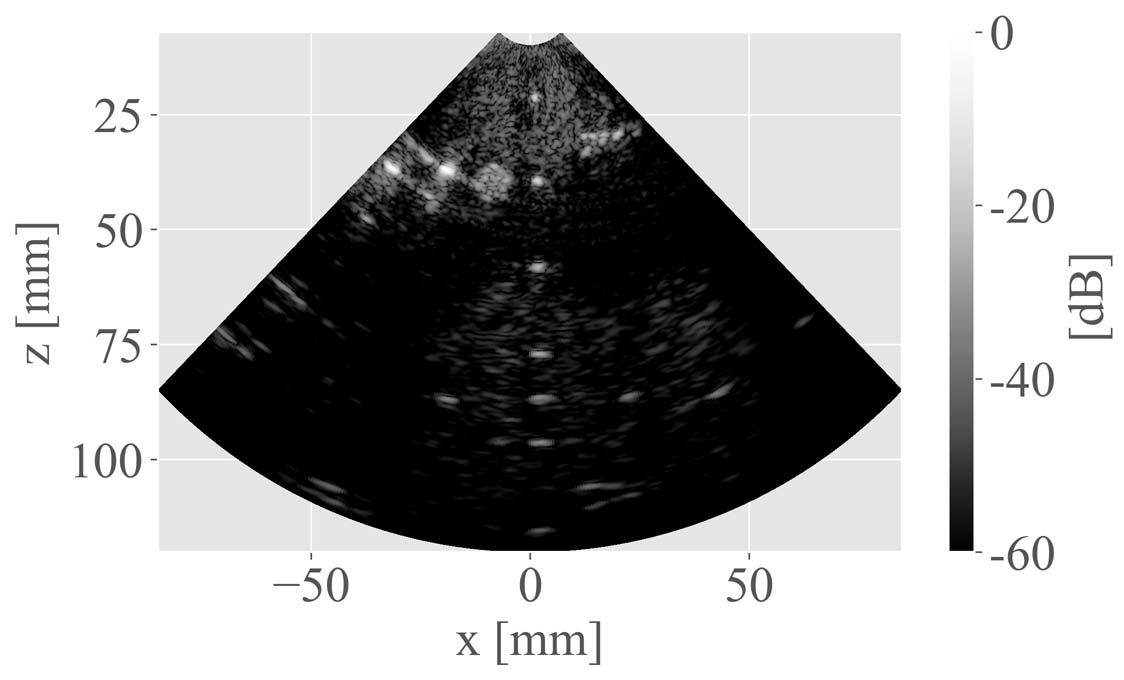}
\label{image3_2}}
\subfloat[]{\includegraphics[width=2.4in]{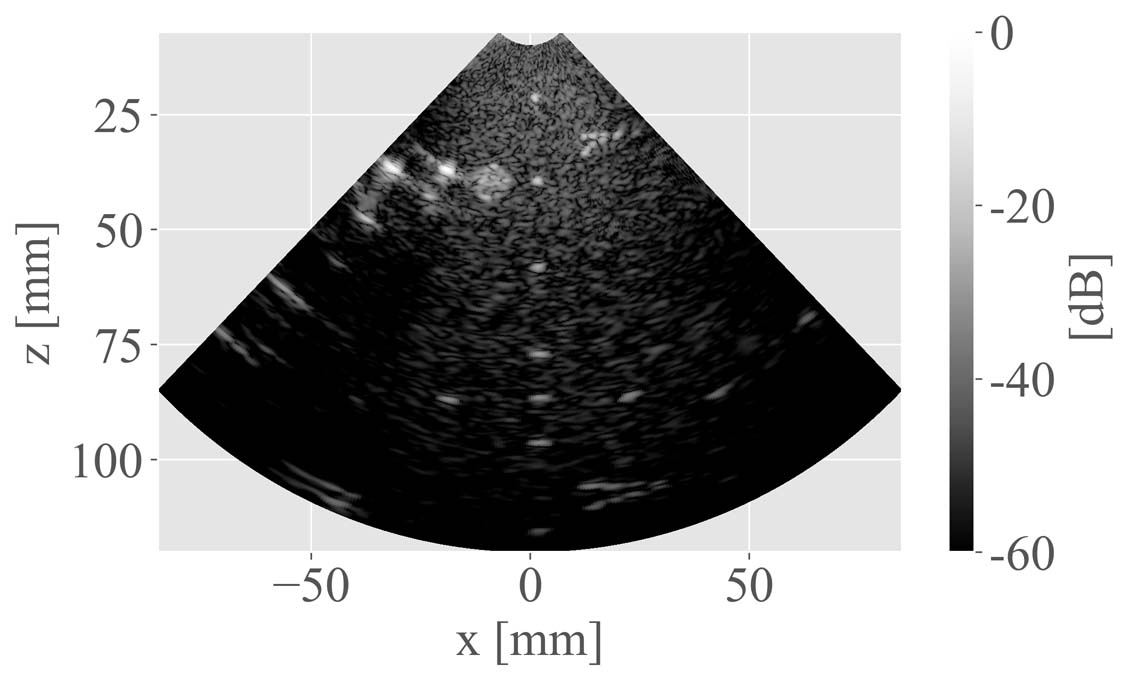}%
\label{image3_3}}

\caption{B-mode images obtained using IDNet and standard compounding. Top to bottom: \textit{in vivo} tissues from the femoris muscle (a, b, c); \textit{in vitro} tissues from the Gammex phantom (d, e, f); and \textit{in vitro} tissues from the CIRS phantom (g, h, i). Left to right: compounding of 3 DWs; reconstruction of IDNet; and compounding of 31 DWs (reference).}
\label{examples}      
\end{figure*}

\section{Results}
\subsection{Performance of the Proposed Network}

In this section, we compare the results of four IDNet models (IDNet-2, IDNet-4, IDNet-8, and IDNet-ReLU) to obtain the optimal performance and verify the contributions of the components of the proposed network. Fig. \ref{crIDNetfigures} displays the images reconstructed from these IDNet models, the standard compounding of 3 DWs, and 31 DWs. Table \ref{inception test} shows the quantitative results reached by the four models. From Fig. \ref{crIDNetrelu}, IDNet-ReLU yielded the worst image quality compared with the models that used maxout activation. It can be observed that there was more amplitude loss in the lateral region. The contrast of the cysts appeared to be decreased; in particular, the cyst in the far field was barely visible. Among the models using the maxout unit, IDNet-2 and IDNet-8 obtained a comparable SSIM and MI compared to IDNet-4, while IDNet-4 performed best in PSNR, CR, CNR, and LR. Therefore IDNet-4 was used as the best IDNet model in the next section, where it was compared with the standard compounding method and other CNN architectures.

\begin{table}[!t]
\renewcommand{\arraystretch}{1.6}
\caption{Evaluation metrics of IDNet and compounding method}
\label{results of IDNet and compounding}
\centering
\begin{tabular}{c c c c}
\hline
model & PSNR [dB] &  SSIM & MI\\ 
\hline
IDNet (3 DWs) & 31.13 $\pm$ 1.47 & 0.93 $\pm$ 0.06 & 0.82 $\pm$ 0.20\\
Compounding (3 DWs)  & 29.24 $\pm$ 1.57  & 0.83 $\pm$ 0.15 & 0.51 $\pm$ 0.16\\
\hline
\end{tabular}
\end{table}

\begin{figure*}[!t]
\centering
\subfloat[]{\includegraphics[width=2.6in]{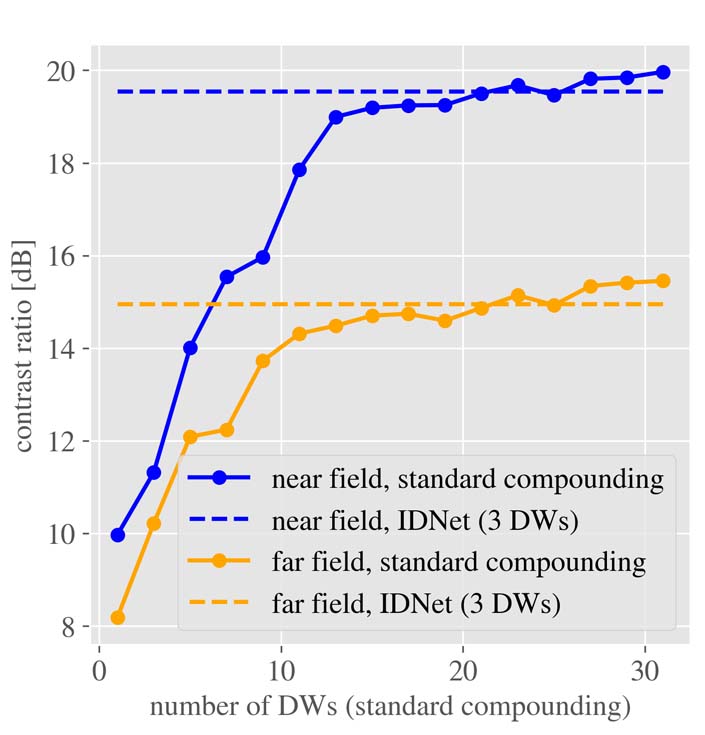}%
\label{cr}}
\hfil
\subfloat[]{\includegraphics[width=2.6in]{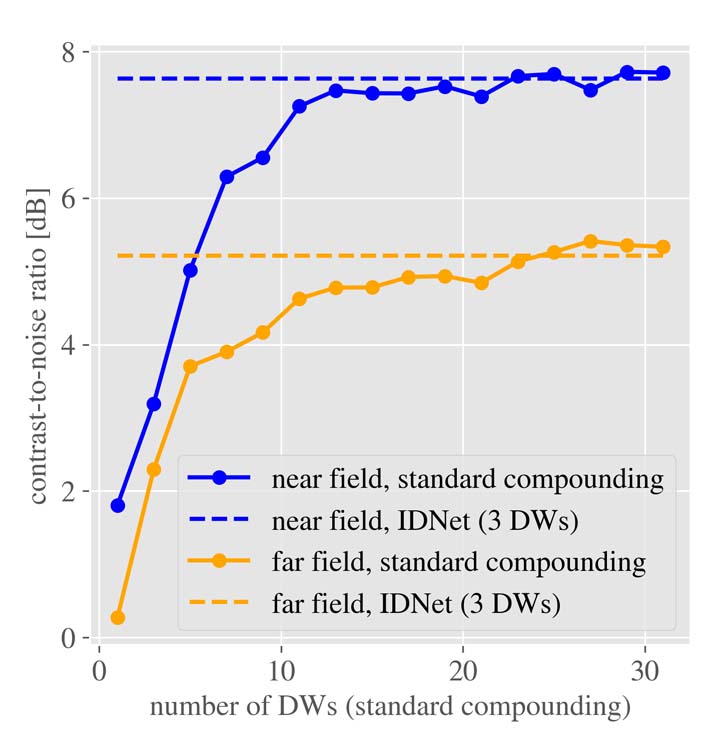}%
\label{cnr}}
\caption{CR (a) and CNR (b) reached by IDNet and standard compounding. IDNet used a constant number of three DWs, whereas the results of standard compounding are given for an increasing number of DWs. The results are given in the near (blue lines and curves) and far (orange line and curves) fields.}
\label{crfigures}
\end{figure*}

\begin{figure}[!t]
\centering
\includegraphics[width=2.6in]{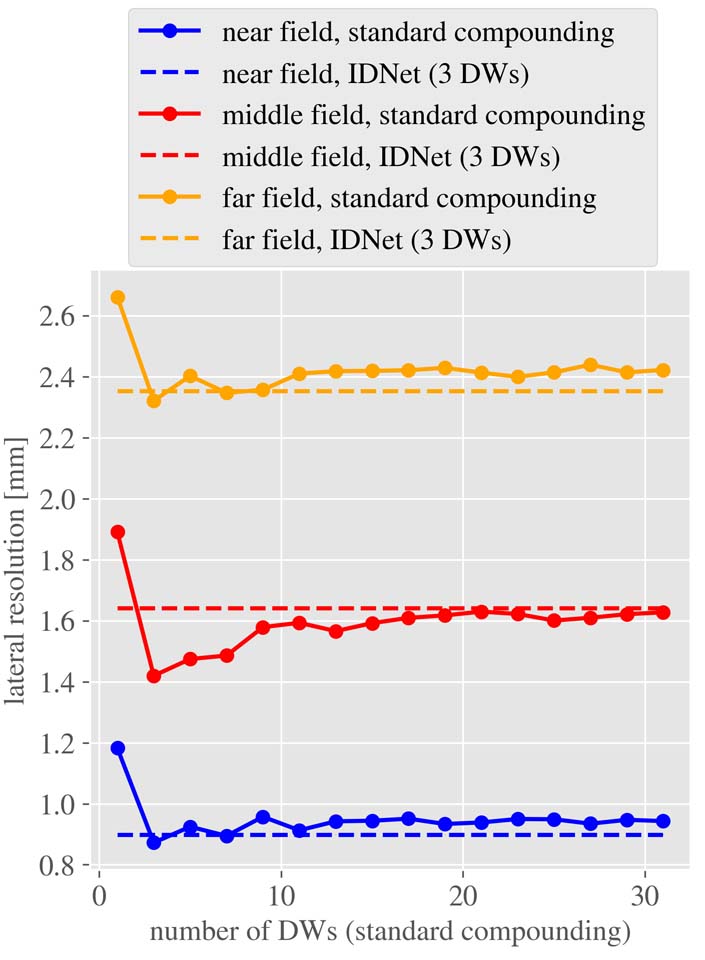}
\caption{LR reached by IDNet and standard compounding. IDNet used a constant number of three DWs, whereas the results of standard compounding are given for an increasing number of DWs. The results are given in the near (blue line and curve), middle (red line and curve), and far (orange line and curve) fields.}
\label{lrfigures}
\end{figure}

\begin{table*}[!t]
\renewcommand{\arraystretch}{1.6}
\caption{Evaluation metrics of IDNet, Gasse's CNN \cite{gasse2017high}, and U-Net \cite{perdios2018deep}}
\label{model test}
\centering
\begin{tabular}{c c c c c c c c c c c}
\hline
\multirow{2} * {model} & \multirow{2} * {PSNR [dB]} &  \multirow{2} * {SSIM} & \multirow{2} * {MI} & \multicolumn{2}{c}{CR [dB]}  & \multicolumn{2}{c}{CNR [dB]} & \multicolumn{3}{c}{LR [mm]} \\ 
\cline{5-11}
~ & ~ & ~&~ & near field & far field & near field & far field & near field & middle field & far field \\
\hline
Gasse et al. & 31.06 $\pm$ 1.46 & 0.93 $\pm$ 0.06 & 0.81 $\pm$ 0.20 & 16.47 & 12.27 & 5.95 & 3.65 & 0.98 & 1.69 & 2.54\\
U-Net        & \textbf{31.15 $\pm$ 1.46} & \textbf{0.94 $\pm$ 0.05} & \textbf{0.82 $\pm$ 0.23} & 14.97 & 12.72 & 5.29 & 4.00 & 1.01 & 1.67 & 2.47\\
IDNet      & 31.13 $\pm$ 1.47 & 0.93 $\pm$ 0.06 & \textbf{0.82 $\pm$ 0.20} & \textbf{19.54} & \textbf{14.95} & \textbf{7.63} & \textbf{5.21} & \textbf{0.90} & \textbf{1.64} & \textbf{2.35} \\
\hline
\end{tabular}
\end{table*}

\subsection{Comparison with compounding method}

The comparison of the visual quality between the reconstruction of IDNet and the coherent DW compounding method is shown in Fig. \ref{examples}. It appears that the quality of the reconstruction of IDNet using only three DWs (Fig. \ref{image1_2}, \ref{image2_2}, and \ref{image3_2}) was improved compared to the images obtained from the standard compounding of the same three DWs (Fig. \ref{image1_1}, \ref{image2_1}, and \ref{image3_1}). In particular, the anatomical structures in Fig. \ref{image1_2} was enhanced compared to Fig. \ref{image1_1}. It can nevertheless be observed that the reconstruction of IDNet appeared to have an amplitude loss in the lateral region (Fig. \ref{image2_2}), as compared to the reference (Fig. \ref{image2_3}).

To quantitatively assess the improvement, we report in Table \ref{results of IDNet and compounding} the PSNR, SSIM, and MI reached by IDNet and standard compounding of 3 DWs, using the compounding of 31 DWs as the reference. The reconstruction of IDNet showed a gain of 1.89 dB in PSNR,  0.1 in SSIM, and 0.31 in MI, as compared to the compounding of the same three DWs. In Fig. \ref{crfigures} and Fig. \ref{lrfigures}, we show the CR, CNR, and LR reached by IDNet and standard compounding. The CR and CNR were measured on the anechoic regions in the phantom (as shown in Fig. \ref{image2_1}, \ref{image2_2} and \ref{image2_3}). The LR was measured on the isolated scatterers (as shown in Fig. \ref{image3_1}, \ref{image3_2} and \ref{image3_3}). IDNet used a constant number of three DWs, whereas the results of standard compounding are given for an increasing number of DWs. From Fig. \ref{crfigures}, it can be observed that IDNet reached a CR and CNR equivalent to those of the compounding of 21 DWs and 23 DWs, respectively, in both near and far fields. In Fig. \ref{lrfigures}, the LR evolution of standard compounding was consistent with the experimental observation from Zhang et al. \cite{zhang2018high}: LR value quickly dropped with two or three DWs and then reached a plateau with more DWs. IDNet reached an LR equivalent to that of the compounding of 3 DWs in the near and far fields, and the compounding of 21 DWs in the middle field.

\subsection{Comparison with other networks}

\begin{figure*}[!t]
\centering
\subfloat[]{\includegraphics[width=2.8in]{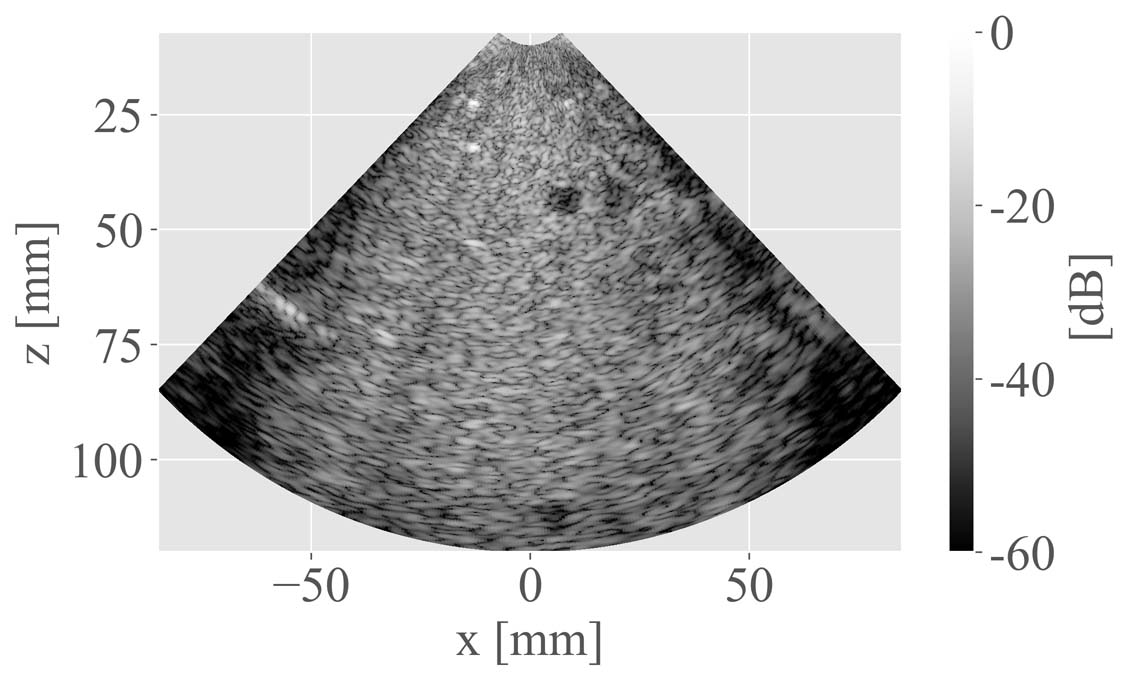}%
\label{crmodel1}}
\hfil
\subfloat[]{\includegraphics[width=2.8in]{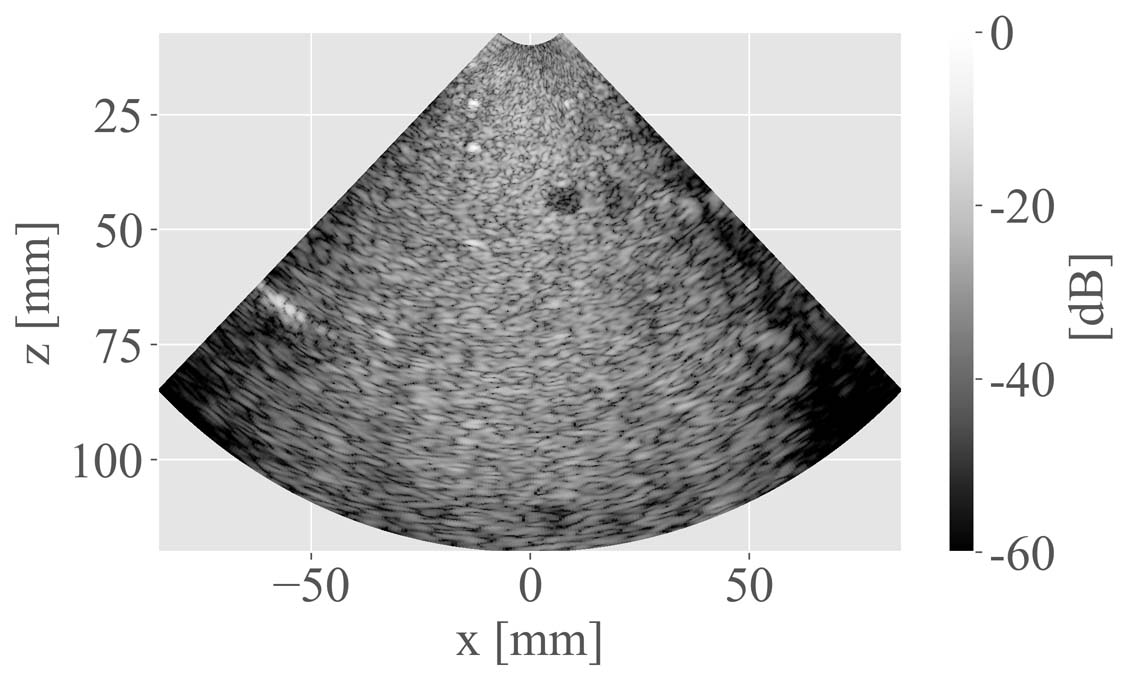}%
\label{crmodel2}}
\hfil
\subfloat[]{\includegraphics[width=2.8in]{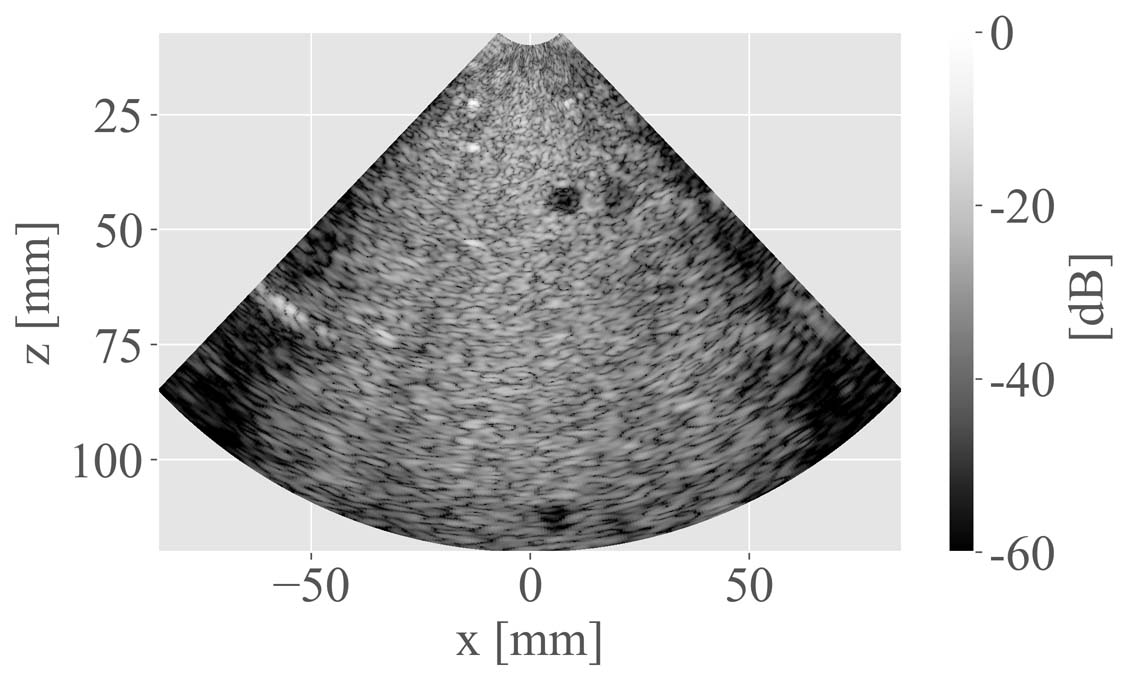}%
\label{crmodel3}}
\hfil
\subfloat[]{\includegraphics[width=2.8in]{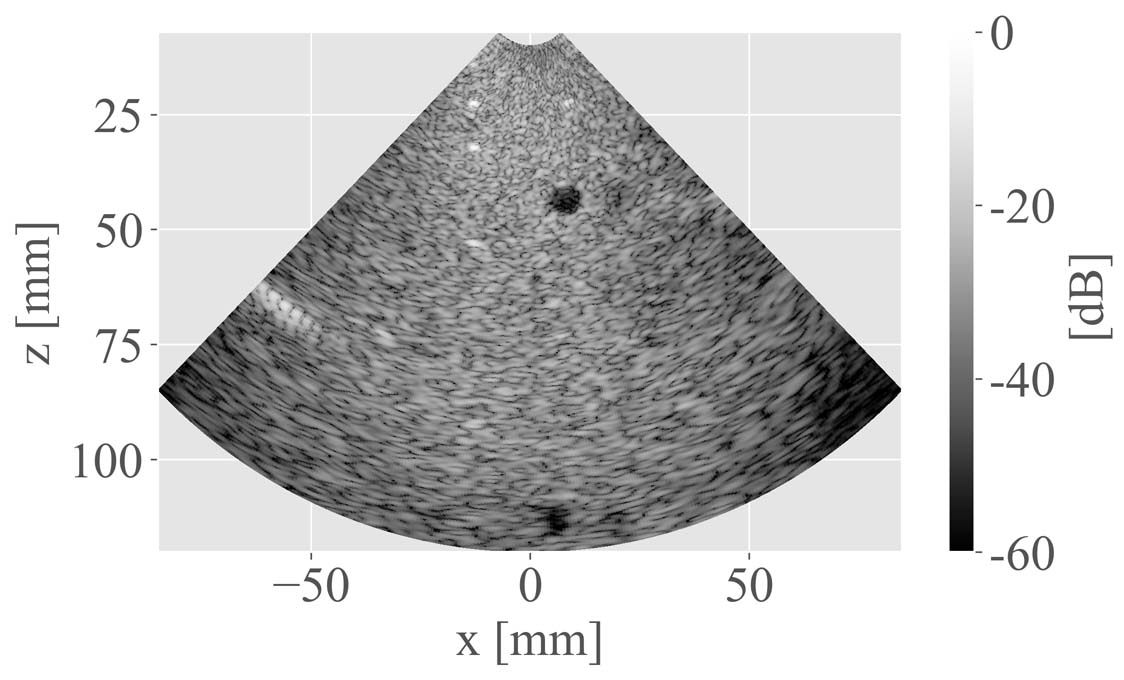}%
\label{crmodel4}}
\caption{Example of B-mode images reconstructed from (a) Gasse's CNN \cite{gasse2017high}, (b) U-Net \cite{perdios2018deep}, (c) IDNet, and (d) standard compounding of 31 DWs (reference).}
\label{crmodelfigures}
\end{figure*}

Using the same input  (DW images of -30$^{\circ}$, 0$^{\circ}$, and 30$^{\circ}$), the reconstructed images from the IDNet, Gasse's CNN, and U-Net are displayed in Fig. \ref{crmodelfigures}. From the figure, it appears that the proposed IDNet produced a better contrast than the other networks, particularly for the cyst in the far field. The quantitative evaluation results of the three models are shown in Table \ref{model test}. The SSIM and MI indices of the three models were rather close, while the PSNR of U-Net was slightly higher than those of IDNet and Gasse's CNN. It can be also observed that the U-Net delivered the worst results in the near field, in terms of CR, CNR, and LR, while producing marginally better values in the middle and far fields, as compared with Gasse's CNN. IDNet produced the best results in terms of CR, CNR, and LR, whatever the depth in the image.

\begin{table}[!t]
\renewcommand{\arraystretch}{1.6}
\caption{Number of parameters, compounding time, and attainable frame rate for IDNet, Gasse's CNN, and U-Net.}
\label{speed test}
\centering
\begin{tabular}{c c c c}
\hline
\multirow{2} * {model} & nb. of parameters & compounding & attainable\\
~ & [million] &  time [ms] & frame rate [fps]\\ 
\hline
Gasse et al. & 1.9 & \textbf{0.48 $\pm$ 0.03} & \textbf{2080}\\
U-Net & 52.7 & 2.38 $\pm$ 0.06 & 420 \\
IDNet & \textbf{1.7} & 0.75 $\pm$ 0.03 & 1330\\
\hline
\end{tabular}
\end{table}

\subsection{Computational complexity and speed}

To evaluate the interest of CNN-based methods in terms of attainable frame rates, all the image formation steps (i.e., including acquisition, beamforming, and compounding) had to be taken into consideration, and to be compared to the steps involved when using standard compounding. Standard compounding implied using 31 DWs, which yielded an acquisition time of 4.96 ms (travel time for 31 DWs for a tissue depth of 120 mm), a beamforming time of 4.46 ms, and a compounding time of 0.02 ms (Please note that for a fair comparison with CNN-based imaging, the computations were performed on GPU). Therefore, the time bottleneck in this case was the acquisition time. The corresponding attainable frame rate was thus 202 fps. The interest of the CNN-based compounding methods was to rely on the acquisition of only 3 DWs, corresponding to an acquisition time of 0.48 ms and a beamforming time of 0.43 ms, so the time bottleneck was the reconstruction time, as shown in Table \ref{speed test}. Table \ref{speed test} gives  for each CNN-based method the number of parameters and the corresponding compounding time (i.e., inference time) and attainable frame rate. Table \ref{speed test} indicates that the U-Net had 52.7 million parameters, which was more than those of IDNet (1.7 million) and Gasse's CNN (1.9 million). In terms of attainable frame rate, the IDNet reached 1330 fps, which was slower than Gasse's network (2080 fps) and faster than the U-Net (420 fps). 

\section{Discussion}

\subsection{High-quality reconstruction for DW imaging using CNN}
In this study, a methodology for the reconstruction of high-quality DW images with deep learning was proposed. We formulated the reconstruction problem as a mapping problem between the low-quality and high-quality images, which was solved by training the proposed CNN architecture (IDNet). Although a large number of samples (a total of 7000 samples used for the training, validation, and testing) and long training time (about two days) were required, once the training was completed, the model could be applied at a high frame rate (1330 fps). The experimental evaluation demonstrated that using 3 DW emissions, the proposed method was able to produce images with comparable quality in terms of contrast and resolution as those obtained from standard compounding of 31 DWs (a ten-fold acceleration factor). It can nevertheless be observed that the deep learning-based methods appeared to yield a loss in amplitude in the lateral area of the images, which could be linked to the fact that not all diverging waves are overlapped in these regions.

\subsection{Inception module to fit sectorial images}

CNNs methods have witnessed a gradual increase in the network depth, corresponding to improvements in various challenges. However, from the experimental results of section IV-C, directly employing conventional CNN architectures for PW imaging (Gasse's CNN using 4 layers or the U-Net using more than 20 layers) yielded a lower performance in terms of contrast and resolution, as shown in Table \ref{model test}. These architectures used convolutions whose shared weights were applied to the entire images or feature maps. This contributed to the shift-invariance property that was applicable in the Cartesian coordinate system associated with PW imaging. Due to the sectorial geometry induced by DW acquisition, using the RF channel data obtained from DW acquisitions as the CNN input implied that the CNN operated in polar coordinates. Thus maintaining the shift-invariance feature for sectorial images required a spatially varying convolution kernel in polar coordinates. To this purpose, we employed the inception module used in conjunction with maxout activation, which contributed to the spatially varying property of the IDNet. In Section IV, the performance of the models using different inception modules and their improvement over conventional networks were investigated. With only two convolution scales in the inception layer, the IDNet-2 model was able to produce equivalent or better results, as compared to Gasse's CNN and U-Net. As the diversity of convolution scales evolved from IDNet-2 to IDNet-4, further improvement of the image quality was obtained. Nevertheless most of the evaluation indices (except the MI and LR in the middle field) of IDNet-8 slightly declined compared with IDNet-4. The performance deterioration of IDNet-8 might be caused by the decreased number of feature maps of each scale in the inception layer. IDNet-4 thus obtained a better compromise between the convolution diversity and feature number of each scale, given a total number of feature maps. 

\begin{figure}[!t]
\centering
\includegraphics[height=1.8in]{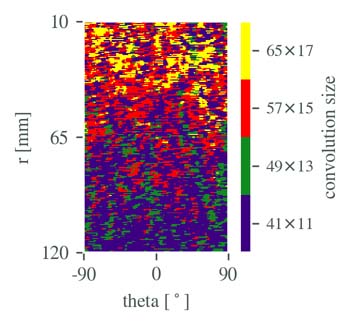}
\caption{Spatial distribution of the inception kernels selected by the last layer of the network. Each pixel indicates which feature map contributed most to generate the output element (i.e., the feature map associated to the maximum coefficient of the 1 $\times$ 1 convolution).}
\label{activation map}
\end{figure}

\begin{figure}[!t]
\centering
\includegraphics[height=1.8in]{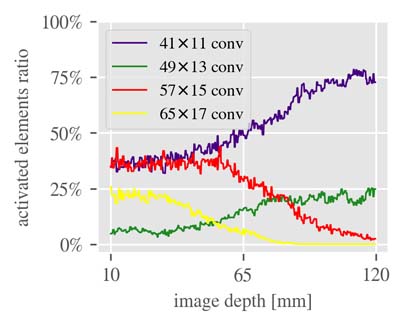}
\caption{Contribution of each inception kernel as a function of depth, expressed as a percentage.}
\label{activation ratio}
\end{figure}

To study the spatially varying property of the proposed method, we analyzed the activation map of IDNet-4. Fig. \ref{activation map} displays the spatial distribution of the inception kernels selected by the last layer of the network (i.e., the 1 $\times$ 1 convolution / maxout layer). 
Each pixel in the figure indicates which feature map contributed most to generate the output element (i.e., the feature map associated to the maximum coefficient of the 1 $\times$ 1 convolution). 
Fig. \ref{activation ratio} further summarizes this distribution by showing the contribution of each kernel for each image depth (expressed as a percentage). It can be observed that the elements generated from the larger convolutions (57 $\times$ 15 and 65 $\times$ 17) tended to contribute more in the near field, while the elements generated from smaller convolutions (41 $\times$ 11 and 49 $\times$ 13) tended to contribute more in the far field, which illustrates the depth-dependence of IDNet. 

%
%

\section{Conclusion}
In this paper, we presented a CNN architecture for the reconstruction of DW imaging. The proposed method aimed at learning a compounding operator to reconstruct high-quality images using a small number of DWs. We demonstrated that the integration of the inception module followed by the maxout activation allowed to exploit information from DW images more effectively. The experimental results demonstrated the effectiveness of the proposed method, yielding an image quality equivalent to that obtained with standard compounding of 31 DWs, which provided a ten-fold increase in the acceleration factor. 

\section*{Acknowledgment}

This work was performed within the framework of the LABEX PRIMES (ANR-11-LABX-0063) of Universite de Lyon, within the program "Investissements d'Avenir" (ANR-11-IDEX-0007) operated by the French National Research Agency (ANR). Financial support from China Scholarship Council (Grant No. 201806120175) to the first author was gratefully acknowledged.

\ifCLASSOPTIONcaptionsoff
  \newpage
\fi

\bibliographystyle{IEEEtran}
\bibliography{DWDeepReconstruction}

\end{document}